RESEARCH ARTICLE                                                                 Open Access

# Seasonality in the migration and establishment of H3N2 Influenza lineages with epidemic growth and decline

Daniel Zinder[1*], Trevor Bedford[2], Edward B Baskerville[3], Robert J Woods[3,4], Manojit Roy[3,5] and Mercedes Pascual[1,3,5]

**Abstract**

**Background:** Influenza A/H3N2 has been circulating in humans since 1968, causing considerable morbidity and mortality. Although H3N2 incidence is highly seasonal, how such seasonality contributes to global phylogeographic migration dynamics has not yet been established. In this study, we incorporate time-varying migration rates in a Bayesian MCMC framework. We focus on migration within China, and to and from North-America as case studies, then expand the analysis to global communities.

**Results:** Incorporating seasonally varying migration rates improves the modeling of migration in our regional case studies, and also in a global context. In our global model, windows of increased immigration map to the seasonal timing of epidemic spread, while windows of increased emigration map to epidemic decline. Seasonal patterns also correlate with the probability that local lineages go extinct and fail to contribute to long term viral evolution, as measured through the trunk of the phylogeny. However, the fraction of the trunk in each community was found to be better determined by its overall human population size.

**Conclusions:** Seasonal migration and rapid turnover within regions is sustained by the invasion of 'fertile epidemic grounds' at the end of older epidemics. Thus, the current emphasis on connectivity, including air-travel, should be complemented with a better understanding of the conditions and timing required for successful establishment. Models which account for migration seasonality will improve our understanding of the seasonal drivers of influenza, enhance epidemiological predictions, and ameliorate vaccine updating by identifying strains that not only escape immunity but also have the seasonal opportunity to establish and spread. Further work is also needed on additional conditions that contribute to the persistence and long term evolution of influenza within the human population, such as spatial heterogeneity with respect to climate and seasonality.

## Background

Seasonal influenza causes considerable morbidity and mortality, and presents a complex problem due to the intimate relationship between its evolution and epidemiology. The WHO estimates influenza A causes between a quarter to half a million deaths worldwide annually [1] with yearly epidemics in the US resulting in tens of thousands of deaths [2]. The economic burden of seasonal influenza in the US is estimated in billions of dollars in health care costs [3,4].

Influenza A is classified into subtypes (e.g. H1N1, H2N2, H3N2) based on its envelope glycoproteins hemagglutinin and neuraminidase, the two major targets of humoral immunity. Multiple zoonotic introductions of influenza A subtypes to the human population have taken place, with H3N2 and H1N1 being the most prevalent subtypes whose continuous endemic circulation has lasted decades.

Sequences sampled since the introduction of H3N2 into the human population in 1968 serve as primary data for phylodynamic inference that seeks to understand joint epidemiological and evolutionary dynamics. H3N2 exhibits rapid geographic spread and turnover rates. On a time scale of several years, all previously circulating lineages are globally replaced by new ones, sharing a

* Correspondence: dzinder@umich.edu
[1]Department of Computational Medicine and Bioinformatics, University of Michigan, Ann-Arbor, MI 48109, USA
Full list of author information is available at the end of the article single and more recent common progenitor (2–8 years) in the past [5-7]. The nature of this swift global turnover remains an open question in terms of both its geographical path and its underlying mechanisms.

It has been proposed that the evolution of the virus is predominantly maintained by a reservoir in the tropics, where annual epidemics experience less severe bottlenecks, which increases the likelihood of local persistence [8]. Further research [9,10] has suggested that the ancestry of global influenza lineages are found mainly in East and Southeast Asia (SEA) rather than the tropics in general, where a network of temporally overlapping epidemics with limited local persistence [11] maintains continuous circulation. An alternative hypothesis suggests that a global metapopulation exists in which temperate lineages frequently revisit the tropics at the end of a seasonal epidemic [9,10]. The much lower contribution of South America (SA) and other subtropical and tropical communities to the long-term evolution of the virus has been attributed to demographics and air travel connectivity [9,12].

Epidemiological and molecular (phylogenetic) studies of influenza use different sources of primary data and their findings are not yet in complete concordance. Both methodologies show support for the common occurrence of strong seasonal epidemics followed by deep troughs limiting viral diversity, for the existence of multiple viral introductions during a season, and for the lack of sustained viral persistence between epidemics [8,13-17]. In contrast, several epidemiological observations, such as spatially structured diffusion patterns [18,19] and hierarchical spread driven by population size and distance, in gravity models [20], have not been evident from phylogenetic methods [16,21,22].

Recently, migration seasonality has been incorporated into phylogenetic analysis, in Bahl et al. [10] where alternative seasonal migration patterns from and to the tropics were considered, and in Bielejec et al. [23] where support for seasonal, rather than constant, H3N2 global migration patterns was established. In Lemey et al. [12], migration rates were assumed to be constant throughout the year, but alternative variables, used as surrogates for measuring the effect of incidence seasonality were tested as predictors of these invariable rates. Specifically, the following features were considered: the overlap in incidence between the source and destination community, the source incidence product with the destination growth rate, and the relative timings of peak incidence. However, these were not informative in predicting these migration rates.

It follows that we still lack an understanding of the relationship between changing incidence throughout the year and the level and timing of immigration and emigration. Specifically, we are interested in whether the timing of migration events with phylogeographic consequence is mainly the outcome of an increased introduction effort from the source (propagule pressure), or is determined by the receptiveness (or the effective R0) of the destination community to introductions. Additionally, we seek quantitative support for the fundamental hypothesis that viral persistence relates to reduced seasonal bottlenecks in incidence.

By using a model that can infer migration rates that differ from season to season, we find clear seasonal migration patterns between and within different global communities. We identify epidemic incidence and growth as predictors of these patterns. Our approach has similarities and differences with [23] on which we expand upon in the Method's section. Notably, we are able to integrate over alternative partitions of the year when measuring the seasonality of migration, and to explicitly incorporate the timing of migration events in addition to migration rates through stochastic mapping.

## Results and discussion

### Seasonal migration to and from, and local persistence within, North America

For this analysis, we are interested in measuring the local persistence of temperate climate North American (NA) H3N2 lineages, and in inferring the seasonal timing of introductions to and from the global community to NA. For this purpose, we partition globally sampled sequences based on their country of collection. Sequences from the US and Canada are designated as NA. A representative sample of "other" global sequences (OT) is randomly sampled in equal proportion from every month in multiple geographic regions. The frequency of sequence sampling (Table 1, Additional file 1: Figure S1) in NA exhibits a winter seasonal pattern, while representative global sequences reflect our equal proportion sampling (Figure 1A). Phylogenetic trees from both the neuraminidase and hemagglutinin proteins are reconstructed based on nucleotide data and sampling time alone. The likelihood of a specific realization of a migration model is marginalized across this posterior distribution of phylogenetic trees in an additional step [24], in which tree likelihood is calculated based on the tree topology, sequence collection locations, and the specified model parameters.

Bayesian variable selection is used to decide whether there is sufficient support for migration between communities, and whether seasonal migration rates, measured in terms of migrations per lineage per unit time, are supported. A

**Table 1 Number of Sampled Sequences in Partition into NA and OT**

| Partition | Hemagglutinin (H3) | Neuraminidase (N2) | Total |
|---|---|---|---|
| NA | 909 | 438 | **1346** |
| OT | 975 | 556 | **1531** |
| Total | **1884** | **994** | **2878** |

**NA** – North America (USA, Canada), **OT** – Representative of global community.

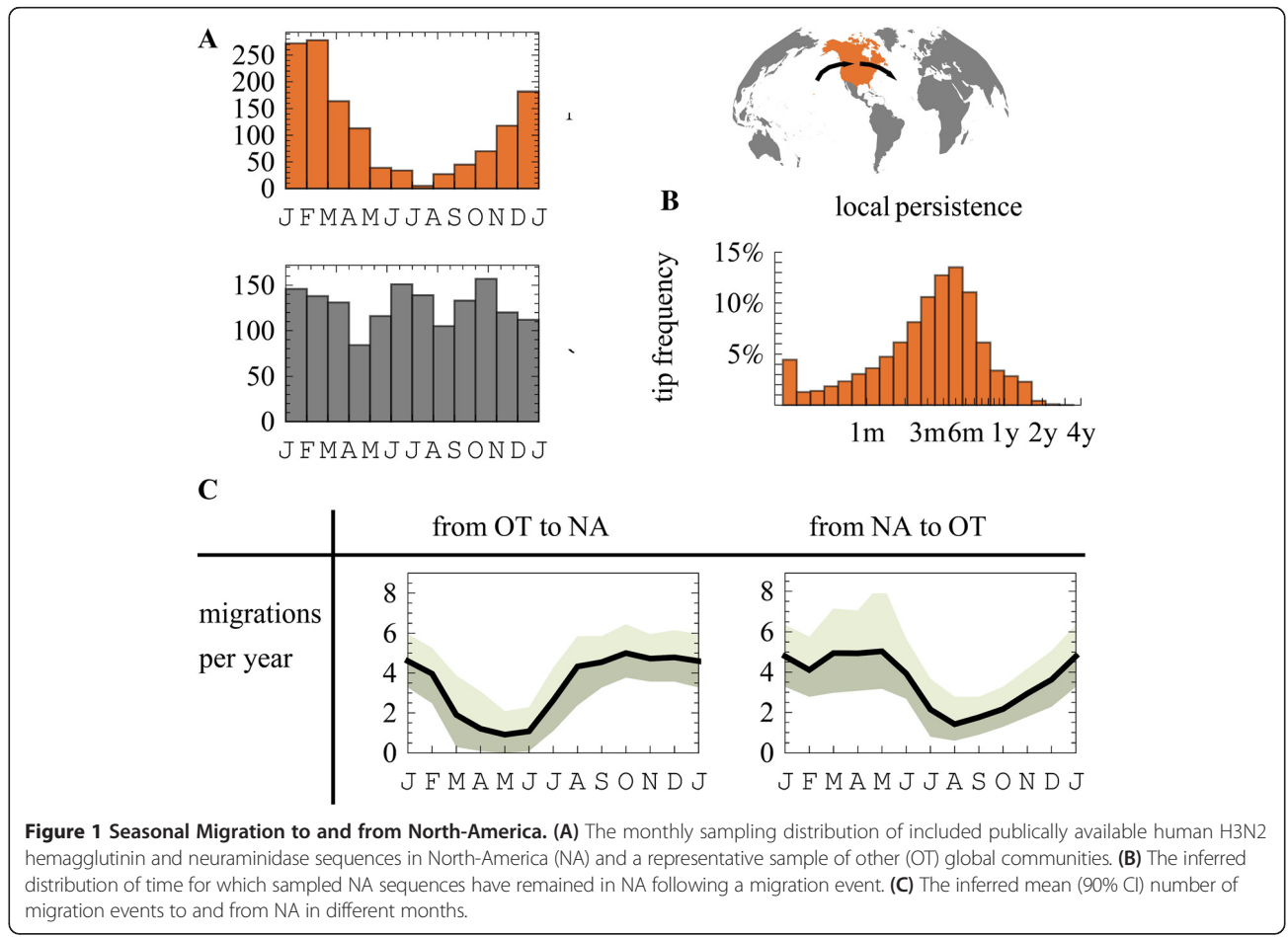

**Figure 1 Seasonal Migration to and from North-America. (A)** The monthly sampling distribution of included publically available human H3N2 hemagglutinin and neuraminidase sequences in North-America (NA) and a representative sample of other (OT) global communities. **(B)** The inferred distribution of time for which sampled NA sequences have remained in NA following a migration event. **(C)** The inferred mean (90% CI) number of migration events to and from NA in different months.

**Table 2 Marginal likelihood of alternative migration models**

| Migration Seasonality | +dof | NA/OT | NC/SC/OT | Global |
|---|---|---|---|---|
| None | 0 | −747.7 | −587.0 | −2246.3 |
| None, variable selection for the presence of any migration between two communities | | | −588.4 | −2241.2 |
| Origin based | n + 1 | | −568.6 | −2193.5 |
| Destination based | n + 1 | | −561.2 | −2207.7 |
| Origin and destination based | 2n + 1 | | −561.1 | −2207.6 |
| Specific origin and destination based | n² − n + 1 | | −561.0 | −2200.6 |
| Specific origin and destination based migration seasonality, variable selection for any migration and for seasonality between each pair of communities | | −726.6 | −561.8 | −2192.6 |

**dof** - degrees of freedom, **NA/OT** – North-America and the Global Community, **NC/SC/OT** – North-China, South China and the Global Community **Global** – 7 + 1 global communities.

seasonal migration model has a higher marginal likelihood than a non-seasonal one (Table 2), and supports migration that is seasonal both to (BF = 70) and from NA (BF > 150). We used this best supported model in further inference.

For each migration model parameterization (Additional file 1: Figure S2) and each tree (see Methods), stochastic mapping is used to sample the internal state of branches and the timing of migration events. Each of these stochastic mappings results in a fully geographically annotated tree sample, including the timing of migration events within branches. Using these data, we can explicitly sample alternative phylogenetic histories from the posterior distribution of trees and model parameters.

Figure 1C shows the mean number of migration events to and from NA in different months. The number of migration events is summed across all lineages, and reported as a yearly rate, during an average month of the year, together with its Bayesian credible interval (referred to as 90% CI). Immigration to NA (Figure 1C) grows during late summer and declines before the end of winter. In contrast, emigration from NA is highest during winter months and during spring. Such emigration from NA during spring could be suggestive of the seeding of epidemics in tropical or southern hemisphere SA [9].

For this and following figures we report the average number of stochastically mapped migration events per unit time in different months, instead of the directly inferred migration rates on a per lineage basis. The number of events per unit time differs from phylogenetic migration rates which take the perspective of a single lineage. The difference is apparent if one considers that a constant migration rate from a specific location measured in terms of a single lineage will necessarily mean many more migration events during an epidemic. We nevertheless report the support (BFs) for seasonal migration, which is used in model selection, with respect to per-lineage phylogenetic migration rates, the unit for which they were originally inferred.

When inferring migration rates we divide the year in two, with constant migration rates inferred individually in each partition. Alternative partitions of the year are weighted according to their likelihood using the MCMC. Each of these partitions and migration model parameterizations is followed by ancestral state reconstruction and subsequent stochastic mapping. The number of migration events per month is counted, and is an estimate of the actual number of migration events expected in the different months, that were captured in our dataset. This can lead to observed positive migration rates during periods with low or zero incidence, such as is the case in NA during its summer incidence trough.

The duration for which tips have been in NA, as traced across the multiple trees is presented in Figure 1B. Sampled NA sequences are inferred to have arrived to NA, 3.8 months (median) before their sampling, with 8% (2-16%) of tips (median and 90% CI) persisting locally for more than one year. Local persistence times for tips are not equivalent to the distance from the trunk (*e.g.* [9,11]). Phylogenetic reconstruction of locations includes stochastic mapping in addition to ancestral state reconstruction (such as in *e.g.* [12]) to resolve branches and uses time variable migration rates which were better supported compared to constant ones.

### Seasonal migration to and from, and local persistence within North China and South China

China is a key source location for H3N2 influenza [9,11]. Here we measure how differing seasonality in northern and southern China relates to H3N2 migration within China and to and from the global community. For this purpose, we use the number of sequences collected in different months to establish broad seasonal patterns in Chinese provinces and several individual cities. For each province, an approximate seasonal pattern is established based on the number of samples in two-months bins, and a clustering algorithm is used to partition the provinces by their incidence seasonality (Additional file 1: Figure S3). This process results in the partitioning of Chinese provinces and associated sequences into a northern (NC) and southern (SC) cluster (Figure 2B). H3 and N2 sequence sampling (Table 3, Additional file 1: Figure S4) in the northern cluster peaks in winter, while the southern cluster sustains two annual peaks (Figure 2A). A representative sample of global sequences was included and designated as (OT). A geographically based partition of China into two was suggested in Du et al. [25] and an alternative partition into three seasonality based clusters was reported in Yu et al. [26]. Although, we find a similar partition to Yu et al. when a division into three clusters is explicitly specified to the algorithm, this partition was not supported over a division into two clusters by our clustering algorithm.

Models that allow for seasonal migration rates between SC and NC and the global community (Table 2) and include seasonality based on the destination community (Additional file 2: Methods 2.7) have the highest marginal likelihood (−561.0 to −561.8) compared to non-seasonal models (marginal likelihood = −587.0, −588.4 without and with variable selection, respectively). Migration seasonality based on the source community has an intermediate marginal likelihood (−568.6). This suggests that both conditions at the source and at the destination community contribute in determining migration rates between the global community, SC and NC. We used a model that includes variable selection for migration and for seasonality between each pair of communities in further analysis (Methods, Additional file 2: Methods 2.17).

We identify strong support (BF > 250 both directions) for migration between SC and NC, and between SC and the global community (BF > 250 both directions). There is support for a model without migration between the global community and NC (BF = 14), and no indication whether migration between NC and the global community (BF = 1.9) is present during the sampled years (Additional file 1: Figure S5). Figure 2D shows the mean number of migration events for different months (black) and their 90% CI across the posterior distribution of model parameters.

Migration events (Figure 2D) from NC to SC peak in January following a peak in incidence in NC and when SC experiences on average a rise in incidence. Migration of SC lineages to NC peaks in October, during epidemic decline in SC and prior to the full onset of the winter epidemic in NC. These patterns are somewhat similar to emigration patterns from NA which remains high during April and May following a decline in incidence. They suggest a potential role for both the propagule pressure (push from the source) and favorable conditions at the destination (fertile ground) in determining effective migration rates.

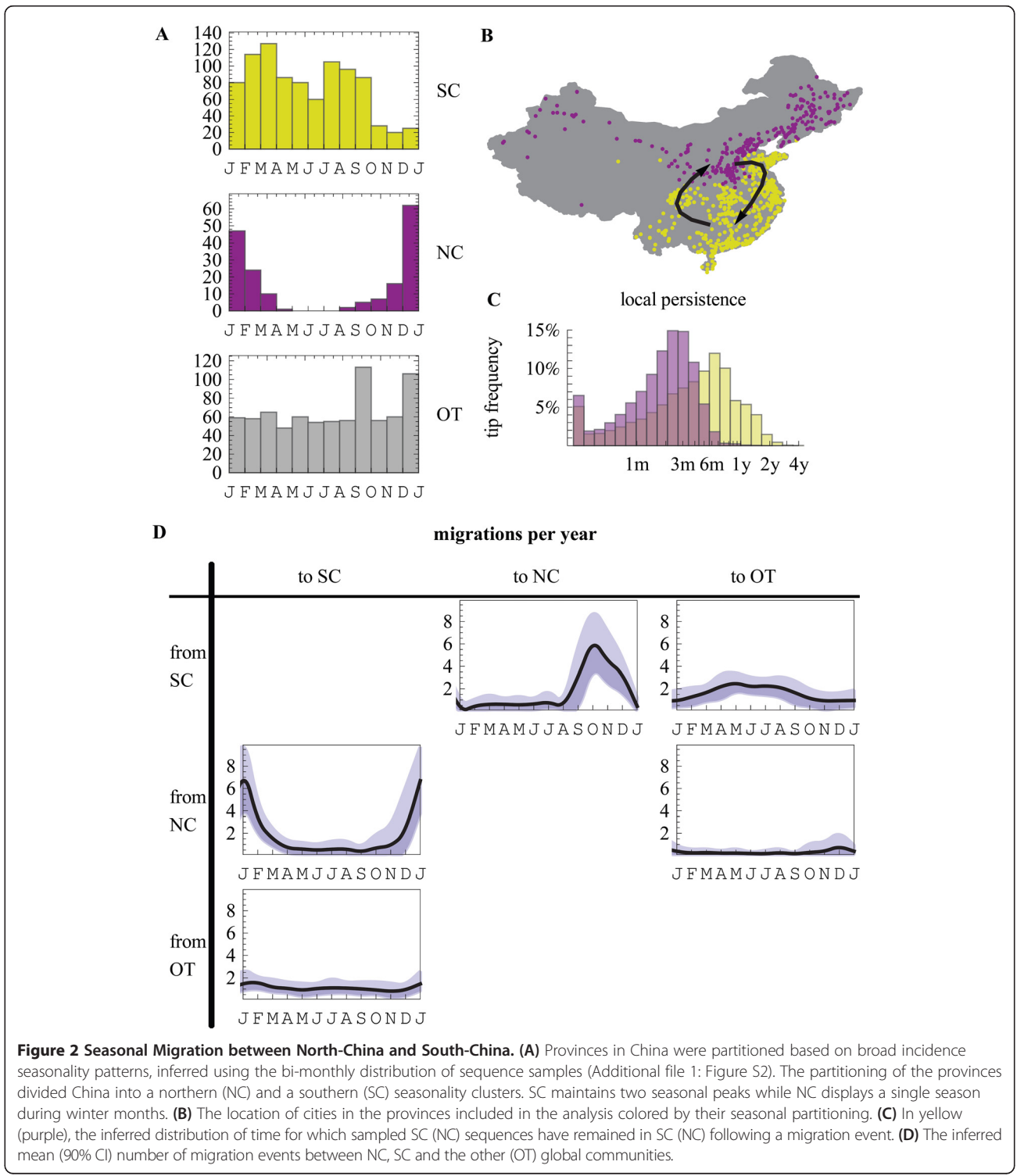

**Figure 2 Seasonal Migration between North-China and South-China. (A)** Provinces in China were partitioned based on broad incidence seasonality patterns, inferred using the bi-monthly distribution of sequence samples (Additional file 1: Figure S2). The partitioning of the provinces divided China into a northern (NC) and a southern (SC) seasonality clusters. SC maintains two seasonal peaks while NC displays a single season during winter months. **(B)** The location of cities in the provinces included in the analysis colored by their seasonal partitioning. **(C)** In yellow (purple), the inferred distribution of time for which sampled SC (NC) sequences have remained in SC (NC) following a migration event. **(D)** The inferred mean (90% CI) number of migration events between NC, SC and the other (OT) global communities.

Local persistence is expected to change with variation in yearly seasonal patterns. We measure the time for which tip ancestry persist locally in SC and NC. Sequences sampled in SC are expected to have emigrated to SC 4.3 months (median) prior to their sampling, where we infer 10% (4-20%) of tips (median and 90% CI) to be locally persistent for more than a year (Figure 2C). NC sequences are inferred to have been in that location for 2 months (median) with 0% (0-2%) of tips (median and 90% CI) locally persistent for more than a year. Since NC

**Table 3 Number of sampled sequences in partition into NC, SC and OT**

| Partition | Hemagglutinin (H3) | Neuraminidase (N2) | Total |
|---|---|---|---|
| SC | 528 | 56 | **584** |
| NC | 150 | 1 | **151** |
| OT | 1302 | 270 | **1572** |
| Total | **1980** | **327** | **2307** |

**SC** – South China, **NC** – North China, **OT** – Representative of global community.

lineages are most often inferred to be derived from recent immigration events, and migration to NC is mainly from SC, we conclude that SC serves as the primary source for NC H3N2 influenza consistent with [25].

Our ability to correctly reconstruct the seasonal timing of migration events depends on the unbiased inference of phylogenetic trees, and the reconstruction of the state of nodes along these branches. Sparse trees reduce our inference power, as branches become uninformative with respect to the underlying migration processes. In addition, the inference of the seasonal timing of migration events is sensitive to the sequence sampling scheme.

### Incidence seasonality and global persistence

To increase the spatial scope of our analyses, we partition both the neuraminidase and hemagglutinin sequences into seven global communities (demes) and an additional representative sample of unclassified sequences (OT) from multiple geographic locations. Sequences were downsampled for computational efficiency, maintaining broad seasonal signals (Figure 3A, Table 4, Additional file 1: Figure S6). A limited (disproportionally higher) number of samples from trough periods were kept to maintain a representation of sequence diversity during troughs. Comparison of alternative, non-seasonal and seasonal models of migration, supported seasonal ones (BF > 33) (Table 2). The best supported model incorporates seasonal migration rates with two partitions of the year, as well as variable selection for the inclusion of migration, and seasonality of this migration, between every pair of demes.

Stochastic mapping is used to infer the state of trunk lineages (2001–2009.5) of both the neuraminidase and hemagglutinin proteins (Additional file 1: Figure S7) taken from the posterior distribution of trees and model parameters. A single stochastically mapped hemagglutinin and neuraminidase tree sample is included in (Additional file 1: Figure S8, S9). We define the trunk of the phylogeny as all the ancestral lineages of the most recent tip samples, discarding lineages which are too young (at most 2.5 years prior to the last tip time).

To show the relation between global persistence and seasonal incidence patterns in different locations, we first generated seasonal incidence profiles from weekly surveillance data (FluNet/WHO who.in/flunet) in each of the seven populations. The yearly incidence of H3N2 changes with age [27], and is expected to have some variation across the populations (e.g. such was the case in A/H1N1pdm09 [28]). However, the mean estimates of the reproduction number are not expected to vary widely [29], and in the absence of available data, we use a simplifying assumption considering yearly attack rates to be similar in each of the seven focal communities. As such, we normalize the incidence seasonality profile based on each community's population size (Figure 3B). We use these profiles to calculate the harmonic mean (HM) of the estimated percent of the total global incidence in each community, in different months, across twelve months. The HM has been used extensively in population genetics when calculating the effects of fluctuations in population size on the effective population size [30] because it captures the increased risk of allelic extinction at low population sizes, *i.e.* the role of population bottlenecks.

We find correlation (N = 14, adj. $r^2$ = 0.41) between the percent of the global trunk inferred to be in a community, and the HM of incidence seasonality. However, when using surveillance data, population size (human population of the countries in a community) was found to be a better predictor (adj. $r^2$ = 0.46, p = 0.01) of the percent of the global trunk in the different demes and was selected for as the only predictor in multiple linear regression model ranking (Figure 4, Table 5) when considered together with the HM and with population density (human population divided by land area).

Although in agreement with the general trend (Figure 4), SA contributes less than expected to the global H3N2 trunk, while NA contributes more. Also, a substantial proportion of the neuraminidase (25%) and hemagglutinin (21%) trunks are inferred to be outside of the seven global communities we sampled (OT).

In other studies, global migration rates were found to be highly correlated with global air travel connectivity [12]. In the case of constant migration rates, in the absence of viral phenotypic evolution, long-term persistence theoretically corresponds to the stationary distribution of the migration rate matrix. However, with the exception of successfully predicting the reduced contribution of SA lineages (Figure 4) [9], this connectivity failed to explain the relative role of different populations to long term viral evolution and therefore, persistence. Seasonal migration rates offer no such single stationary distribution, but may result instead in a periodic cycle between alternative attractor configurations throughout the year (in the absence of phenotypic evolution). As such they have the potential to better describe a global metapopulatoin structure in which the trunk of the phylogenetic tree travels throughout the year.

Since the same predictor values are used to infer both the neuraminidase and the hemagglutinin trunk proportions,

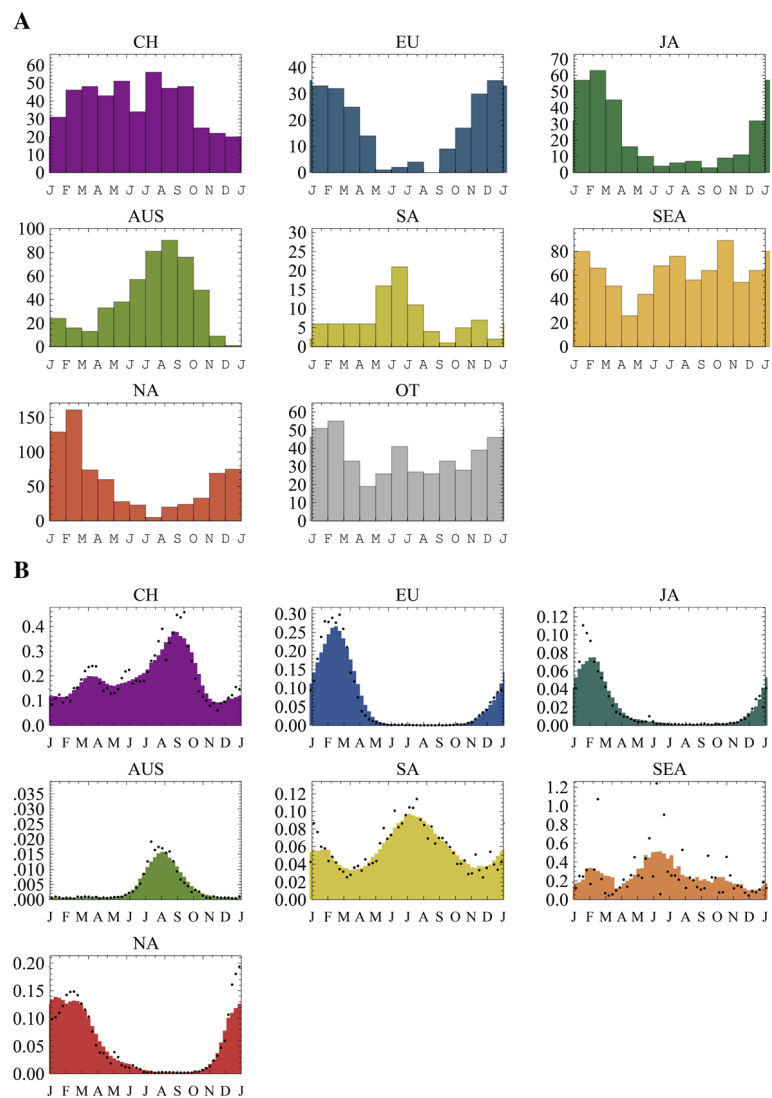

**Figure 3 Sequence Sampling and Incidence Profiles for Seven Global Communities. (A)** The monthly sampling distribution of global (CH – China, EU – Europe, JA – Japan, AUS – Oceania, SA – South-America, SEA – South East Asia, NA – North-America) publically available human H3N2 hemagglutinin and neuraminidase sequence samples used for the inference of phylogenetic trees and migration seasonality (Table 4, Additional file 1: Figure S4). A representative collection of sequences sampled from other parts of the world is designated as other (OT). **(B)** Surveillance data (WHO FluNet, 2000 week 1 to 2012 week 52) was aggregated on a weekly basis and smoothed (8 weeks moving average) to obtain broad seasonal incidence profiles in seven global communities. Within each global community, country level surveillance counts were normalized (divided by the total number of counts in the country), and added up in proportion to the country's population size. Each community was weighted based on its population size to approximate its relative contribution to worldwide incidence.

and we are limited to seven predictor values, our results are sensitive to variation in these predictors such as the underlying seasonal patterns or the estimates of population sizes.

Estimates, of the percent of the global trunk in each community, are dependent on the sampling and inclusion of near trunk lineages in the inference, the discovery of which is dependent on the sampling effort and on incidence itself, both of which vary from location to location, and with time. Our results largely correspond to the Bayesian credible intervals reported in [9], in which sampling considerations were taken into account.

An important factor which was not considered in our analysis is the role of spatial patterns of incidence on viral persistence and long term evolution. Increased or reduced spatial correlation with respect to seasonality across large geographic regions, may account for additional unexplained variability between the regions with respect to the amount of long term viral evolution that they sustain. This is the case as fadeouts in incidence were observed in individual countries in, e.g. SEA [11], while our analysis of seasonal incidence patterns aggregates these countries together. In the future, better availability

Table 4 The number of sequences used in seven global communities

| Partition | Hemagglutinin (H3) | Neuraminidase (N2) | Total |
|---|---|---|---|
| CH | 320 | 151 | **471** |
| EU | 134 | 69 | **203** |
| JA | 180 | 83 | **263** |
| AUS | 172 | 314 | **486** |
| SA | 86 | 5 | **91** |
| SEA | 309 | 429 | **738** |
| NA | 288 | 413 | **701** |
| OT | 236 | 188 | **424** |
| Total | **1725** | **1652** | **3377** |

CH – China, EU – Europe, JA – Japan, AUS – Oceania, SA – South-America, SEA – South East Asia, NA – North America, OT – Representative of unclassified sequences from multiple geographic locations.

of data from extensive year-round sequence sampling may allow more detailed partitioning of the world population and could help mitigate some of these limitations, as well as improve our estimation of the contribution of different mechanisms in maintaining long-term viral evolution.

### Incidence seasonality and global migration

Multiple mechanisms contribute to the number of effective migrations, that is, to migration events that are not lost to rapid extinction and are able to achieve a sufficient population size to be picked up in a sequencing study that is deposited in the database. Contact between the source and destination community may depend on the pathogen abundance at the source, and is also modulated by factors such as air travel. Furthermore, once at its destination, an invading pathogen experiences different seasonal transmission rates and host susceptibility levels. This is further complicated by the dependence of the invasion on antigenic or other traits of the invading pathogen [12,31].

Here we wish to establish how two basic features of seasonal epidemics, namely incidence and growth, correlate with the number of emigration and immigration events observed along the H3N2 phylogeny. An understanding of these associations provides a basis for identifying the processes contributing to the seasonality of global H3N2 migration.

As before, we use surveillance data from FluNet/WHO to obtain a broad representation of the seasonal patterns in each community (Figure 3B). Whereas in the above analyses the more likely partitions of the year were selected by the MCMC, here, each partition of the year is centered along a different month in order to prevent the confounding of the identification of a likely partition with our measurements of incidence and growth during the same six month period. For each partition, we use our MCMC framework to infer migration rates between each pair of communities. This is done jointly for tree pairs from the sampled distribution of hemagglutinin and neuraminidase trees. We use stochastic mapping to map emigration and immigration events along branches of the phylogeny (Additional file 1: Figure S10).

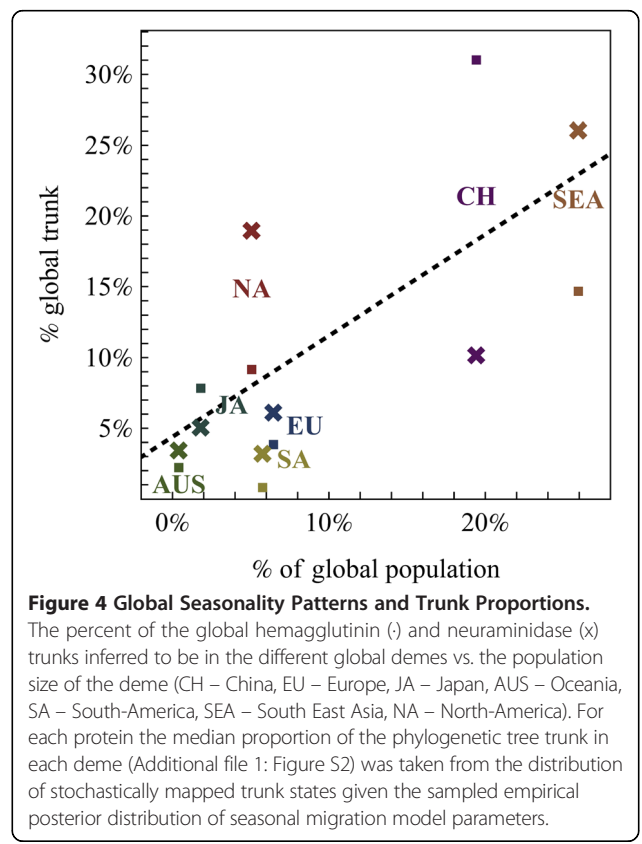

**Figure 4 Global Seasonality Patterns and Trunk Proportions.** The percent of the global hemagglutinin (·) and neuraminidase (x) trunks inferred to be in the different global demes vs. the population size of the deme (CH – China, EU – Europe, JA – Japan, AUS – Oceania, SA – South-America, SEA – South East Asia, NA – North-America). For each protein the median proportion of the phylogenetic tree trunk in each deme (Additional file 1: Figure S2) was taken from the distribution of stochastically mapped trunk states given the sampled empirical posterior distribution of seasonal migration model parameters.

For each of seven global communities and six alternative partitions of the year into two equal parts, we measure the fraction of immigration events to, or emigration events from, the community (N = 42) during the corresponding six months. This is repeated across multiple

Table 5 Ranking of alternative linear regression predictors for the percent of the global H3N2 trunk in different communities

| Model | Adjusted $R^2$ | $R^2$ | AIC | BIC |
|---|---|---|---|---|
| {Pop} | 0.464411 | 0.506 | 226.15 | 228.06 |
| {HM, pop} | 0.421031 | 0.510 | 228.02 | 230.57 |
| {Pop, density} | 0.419179 | 0.508 | 228.06 | 230.62 |
| {HM} | 0.414195 | 0.459 | 227.40 | 229.32 |
| {HM, pop, density} | 0.367029 | 0.513 | 229.93 | 233.13 |
| {HM, density} | 0.365295 | 0.463 | 229.30 | 231.86 |
| {} | 0. | 0. | 234.00 | 235.29 |
| {Density} | −0.0538263 | 0.027 | 235.62 | 237.53 |

**HM** – Harmonic mean of the estimated percent of the total global incidence in each community, in different months, across twelve months **Pop** – Population size of a community, as total of the countries included in a community, **Density** – Population density of a community, as the total area divided by the total population size of countries within the community.

samples from the posterior distribution of rate parameters. We then apply multiple linear regression models to identify possible correlations between incidence and growth to median immigration and emigration. The fraction of positive growth in a six month window is calculated as the fraction of months in that period that show an increase in incidence compared to the previous month based on yearly incidence profiles from surveillance data.

Immigration is significantly correlated (adj. $r^2 = 0.69$) with both absolute incidence (adj. $r^2 = 0.22$, $p = 2\times10^{-6}$) and positive growth (adj. $r^2 = 0.52$, $p = 6\times10^{-9}$) (Figure 5C, D). There is also correlation (adj. $r^2 = 0.13$, $p = 0.01$) between epidemic decline and increased emigration (Figure 5B). We find however no support for correlation between the fraction of the yearly incidence within a single community during a six month long period, and the fraction of emigration during that same period (Figure 5A). The global migration patterns largely correspond to the two case studies, where the absolute incidence as well as growth in incidence, are indicative of higher immigration, while epidemic decline is associated with increased emigration (Figure 5).

These results correspond to a case where sufficient propagule pressure is achieved at varying levels of incidence at the source. Thus, the amount of effective emigration does not correlate directly with incidence. This may be the case when sufficient contact between the global communities is reached before and after peak incidence levels. In contrast, the results suggest a much greater role for conditions suitable for growth (fertile grounds) at the destination, in determining the probability that an immigration event will lead to successful establishment and be counted as an effective immigration event.

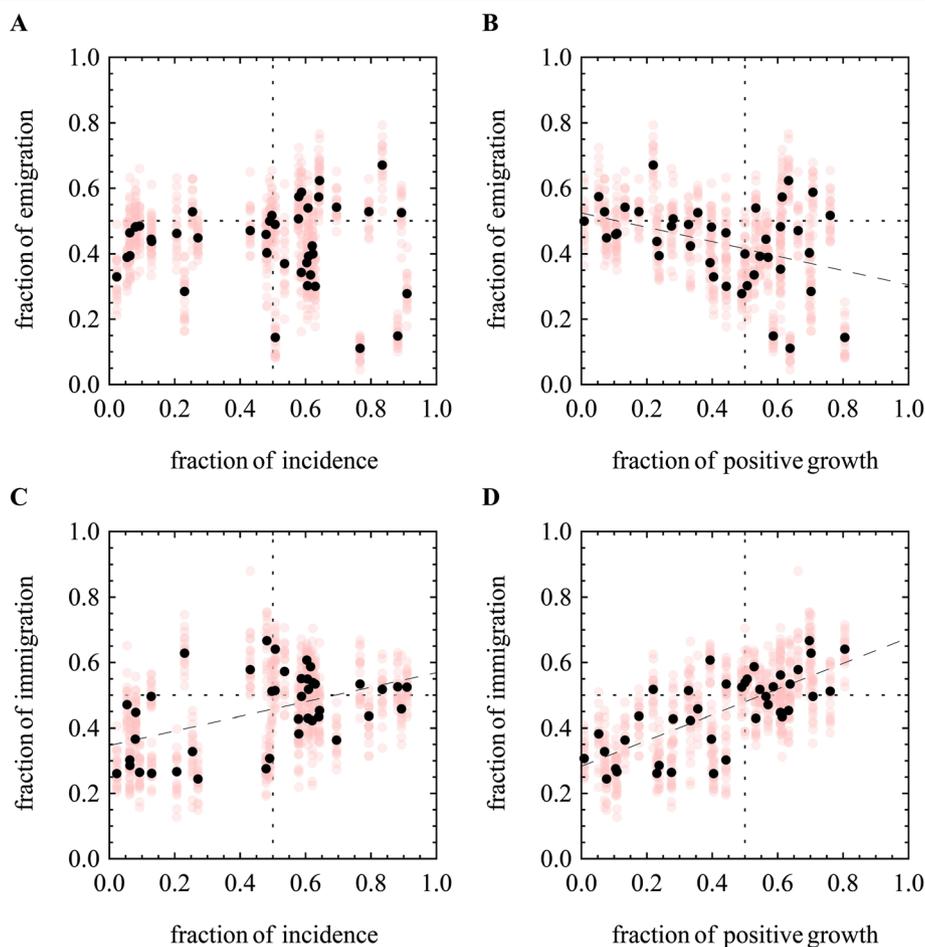

Figure 5 Correlation between Growth, Incidence, Immigration and Emigration. Seasonal incidence profiles (Figure 3B) in each community, are used to identify gross periods of growth, decline, and to estimate the monthly % of the yearly incidence during six month periods, averaged across multiple years. Alternative seasonal migration models, partitioning the year into two, and centered on consecutive months (Jan-May) were parameterized using the MCMC. The number of stochastically mapped migration events between each pair of locations was counted in each of the six month long partitions. (A) Medians (black) and samples (gray) of the fraction of the total emigration events from a location during a six month long period, vs. the fraction of the yearly incidence in the corresponding six month period in the source location. (B) The fraction of the total emigration events from a location during a six month long period, vs. the fraction of the yearly positive growth in the corresponding six month period in the source location. (C) The fraction of the total immigration events to a location during a six month long period, vs. the fraction of the yearly incidence in the corresponding six month period in the destination location. (D) The fraction of the total immigration events to a location during a six month long period, vs. the fraction of the yearly positive growth in the corresponding six month period in the destination location.

### Limitations

Bayesian inference and the estimation of marginal likelihoods are dependent on the choice of priors. These serve as baseline assumptions for the model parameters, and sufficient data to the contrary will pull the estimated parameters away from these assumptions. We used constant-population coalescent process priors because they are well established in the context of Influenza A phylogentic inference [9,10,32-34]. These priors are simpler to implement, and had little influence on phylogeographic inference in previous reports [32]. Therefore alternative tree priors, such as ones involving birth death processes [35], were not used.

In addition, the inference of trees based on combined sequence data and geography, rather than sequential consideration of these two aspects as implemented here, should be more accurate by allowing the more thorough exploration of tree space. Although in general this scheme will sample tree space more exhaustively, it will incur a computational effort which will at the least amount to multiplying the computational effort by the number of models evaluated. As most of the information in our analysis (magnitude of 10,000s vs. 1,000s log likelihood units) relating to tree topology is contained in nucleotide data compared to geography, we expect the tree samples to be sufficiently representative, and the combined inference of trees and migration processes to be equivalent to a sequential one. Some of these considerations are listed in [24].

The availability of sequence samples is limited, and we assume that a sufficiently representative sample of sequences is available from each community and across time. Sufficient sampling is required in order to generate dense trees which will contain information about the correct seasonal timing of migration events. Differences in sampling and sequence availability may bias the amount of migration inferred to take place between two communities. When analyzing the results, we refer only to the seasonality of migration and not to the total amount of migration throughout the year. Although we expect this quantity to be less sensitive to sampling, it may be that an increase in the sampling of one community vs. the others will bias the timing of immigration to weigh more towards the timing relevant to this community. We acknowledge that this is a limitation of our analysis which cannot be completely avoided at the moment as the number of available samples globally is highly skewed towards specific countries, and homogenous sampling reduces the power of our analysis considerably. In the future with the increase in worldwide sampling cover, alternative sampling scenarios could be considered. In agent based simulations [7] we evaluated similar sampling scenarios as the ones used here, and found the results of the inference largely consistent with the underlying migration processes (Additional file 2: Methods 3.3).

### Conclusions

Our results show clear support for seasonal variation in migration rates. We used models incorporating this variation to estimate patterns of global seasonal migration and of persistence.

H3N2 persistence is short, on the scale of several months, with only a small fraction of the lineages persisting for over a year (Figures 1B and 2C, Additional file 1: Figure S11). Seasonal patterns also correlate with the probability that local lineages go extinct and fail to contribute to long term viral evolution. However, the probability that a region will contribute to long term viral evolution as a part of the trunk of the phylogenic tree was found to be better determined by its overall human population size. In general, this short local persistence indicates a massive replacement of circulating lineages on both annual and sub-annual timescales, much shorter than those characteristic of global turnover, typically around several years.

This rapid replacement is mediated by migration which is by itself highly seasonal in nature (Table 2). In particular, the likelihood of successful immigration increases during periods of the year that coincide with epidemic growth and higher incidence, a pattern suggestive of a 'fertile ground' hypothesis, where incoming viruses survive and spread more effectively during this upward season (Figure 5C,D). Emigration, on the other hand, coincides with periods of epidemic decline (Figure 5B), which suggests a 'tail-to-beginning' migration pattern between overlapping epidemics. These overlapping seasons would correspond to the major epidemics of the two hemispheres which are known to exhibit opposing seasonality. This pattern could also indicate the reintroduction of lineages to the tropics at the end of temperate climate epidemics as was suggested by Bedford et al. in [9].

Surprisingly, the dominant factors behind influenza's incidence seasonality remain a subject of debate, with emphasis on either environmental factors influencing transmission or host susceptibility [36-40]. With respect to migration, focus has been largely given to factors such as connectivity through air travel, and not to other aspects of effective migration related to conditions at the source and destination. The phylogenetic tree of the virus, if correctly reconstructed, can provide multiple natural experiments involving the state of the environment and the phylogenetic outcome. Incorporating seasonality into the reconstruction of environmental conditions appears essential.

## Methods
### SeasMig

We implemented in Java a tool (http://bitbucket.org/pascualgroup/seasmig). A detailed description of *SeasMig* in the context of this manuscript is available in Supplementary Methods. Using *SeasMig* alternative migration models parameters could be inferred and compared by their marginal likelihood including seasonal, epochal, and non-seasonal phylogeographic migration models. An empirical distribution of trees in nexus format (*e.g.* generated using BEAST [41]) is given as input. Our tool uses an MCMC to sample from the posterior distribution of model parameters and stochastically mapped migration events along branches and trunk lineages (Figure 6). Multiple MCMC chains run in parallel and perform chain swaps in a lockstep manner.

### Stochastic mapping

Stochastic mapping is an additional step following the calculation of tree likelihood and ancestral state reconstruction at the nodes of a tree. This mapping allows us to generate a stochastic realization of the state of branches along the tree, in addition to the state of internal nodes, and in so doing, provides samples of migration and mutation events, and their timing along the tree that lead to the observed tip states. Stochastic mapping of both sequence (nucleotide) and character (*e.g.* geographic) annotations is available in *SeasMig*, together with the option of incorporating seasonal migration models. Stochastic mapping is implemented directly in our code based on [42]. Improved performance could be achieved using [43]. Briefly, a given type of event, migration or mutation, is assumed to behave as a Poisson process along a branch. As such, the timing of the next event follows an exponential distribution with a mean equal to one over the total rate of emigration (mutation from) a location (a base). Once the timing of the next event is determined, the event is chosen based on its relative probability compared to other emigration events. Branch reconstructions that span different seasons were performed by sampling the state (here the location) within the first season's boundaries using the initial migration matrix, and by continuing the stochastic mapping forward using the second matrix.

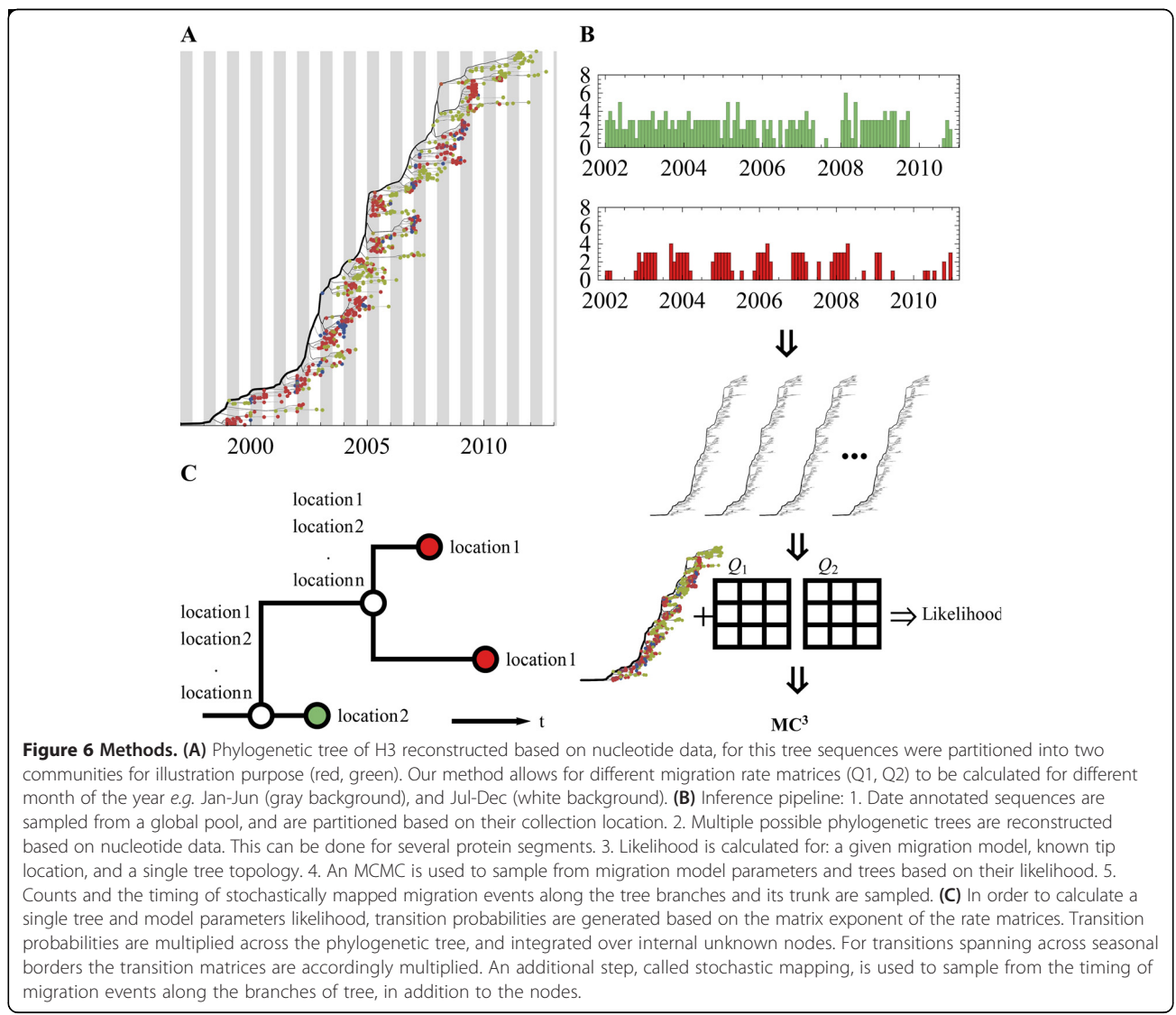

**Figure 6 Methods. (A)** Phylogenetic tree of H3 reconstructed based on nucleotide data, for this tree sequences were partitioned into two communities for illustration purpose (red, green). Our method allows for different migration rate matrices ($Q_1$, $Q_2$) to be calculated for different month of the year *e.g.* Jan-Jun (gray background), and Jul-Dec (white background). **(B)** Inference pipeline: 1. Date annotated sequences are sampled from a global pool, and are partitioned based on their collection location. 2. Multiple possible phylogenetic trees are reconstructed based on nucleotide data. This can be done for several protein segments. 3. Likelihood is calculated for: a given migration model, known tip location, and a single tree topology. 4. An MCMC is used to sample from migration model parameters and trees based on their likelihood. 5. Counts and the timing of stochastically mapped migration events along the tree branches and its trunk are sampled. **(C)** In order to calculate a single tree and model parameters likelihood, transition probabilities are generated based on the matrix exponent of the rate matrices. Transition probabilities are multiplied across the phylogenetic tree, and integrated over internal unknown nodes. For transitions spanning across seasonal borders the transition matrices are accordingly multiplied. An additional step, called stochastic mapping, is used to sample from the timing of migration events along the branches of tree, in addition to the nodes.

### Tree sampling

For each of the inferences, we use a sample of trees taken from the empirical posterior distribution of trees generated by BEAST based on sequence data and sampling times alone. Sampled sequences are aligned using MUSCLE [44], sequences with low alignment quality were manually removed. We perform phylogenetic tree reconstruction of coding region sequences with high coverage using BEAST 1.7.4. A differential codon location evolutionary model is used (HKY1 + 2) [45]. Models with a biological meaning, which account for variation in the evolutionary rate in different codon positions, such as the HKY1 + 2, were found to be better supported than standard nucleotide substitution models such as general time reversible with gamma distributed rate heterogeneity and a proportion of invariant sites (GTR + Γ + I) in most RNA viruses in [45] while requiring less parameters. In each analysis 2000 trees from the stationary distribution of four independent chains sampled every 10000 steps, are combined. BEAST XML, tree and log files are available through Dryad [46]. BEAST tree inference was carried out on computational resources and services provided by Advanced Research Computing at the University of Michigan.

### Sequence sampling

We sampled sequences from the NCBI flu database. All, but at most n samples chosen randomly per k consecutive months per community are used. This sampling scheme is intended to reduce the number of overall samples for computational reasons, by decreasing the number of samples taken in more recent years as more sequences were generated, while maintaining available data during the seasonal troughs. For the sampling of "other" global sequences (OT), equal proportion sampling in every month for every geographic location was used with the objective of capturing a sufficient representation of the underlying global genetic diversity. The number of sequences used in each analysis is included in Tables 1, 3 and 4 and in Additional file 1: Figures S1, S4 and S6. Sequences were sampled between 1999 and 2013; their accession numbers, dates, and geographic classification, for the different figures are included in Additional file 3. In simulation studies, the seasonal pattern of effective migration events was better reconstructed when tip (sequence) sampling was proportional to incidence (Additional file 2: Methods 3.3.2.2), as opposed to when sampling was uniform across time and in different populations (Additional file 2: Methods 3.3.2.1).

### Seasonal migration model

We generate a seasonal migration model (Figure 6) by using two different constant migration rate matrices ($Q_A$ and $Q_B$) for two parts of the year labeled as "season A" and "season B" respectively. To estimate the transition probabilities between two geographic locations at different times, we calculate the respective transition probability matrices P for the individual constant rate periods through matrix exponentiation [23]. For the complete time interval, the individual transition probability matrices are multiplied accordingly. Given a tree topology, we integrate unknown internal node states over the tree efficiently by caching conditional probabilities of individual node states as described in [47]. Our method has similarities with [23] but is parameterized differently as detailed in the next paragraph. This parameterization allows a smooth transition from a seasonal to a non-seasonal model which is well suited for variable selection. Our approach is also different because it considers alternative partitions of the year which are either sampled using the MCMC in proportion to their likelihood (in Figures 1, 2 and 4) or are integrated upon with equal probability (in Figure 5). The latter approach allows us to consider correlates of migration seasonality realted to incidence independent of the choice of the likely partitions of the year.

### Seasonal migration model parameterization and priors

We parameterize migration rates for the two partitions of the year as $r_{from,\ to} \cdot (1 + \sigma_{from,\ to})$ and $r_{from,\ to} \cdot (1 - \sigma_{from,\ to})$. This is done with the purpose of measuring the inferred rates, and their seasonality separately and allowing for separate indicators, used in variable selection, for the inclusion or exclusion of seasonality, and for the inclusion and exclusion of any migration from one community to another. Migration rates $r_{from,\ to}$ are drawn from an exponential prior, giving a non-diminishing probability to high migration rates. The 'seasonal scaling' $\sigma_{from,\ to}$ parameter gives the relative increase (and decrease) in contribution of migration in one season compared to the average migration rate. This parameter is sampled from a uniform U(−1,1) prior.

### Metropolis-coupled MCMC (MC3)

We use an MC3 algorithm to sample model parameters and to from the sample of trees. Metropolis-coupled Markov-chain Monte Carlo, or MC3, is an MCMC algorithm that allows sampling from analytically intractable distributions, and builds on standard MCMC by improving mixing [48]. Such distributions include the distribution of tree likelihoods given a mutational or a migration model. In particular, MC3 includes multiple MCMC chains: a cold chain samples from the target distribution, while hot chains sample from a flattened likelihood surface exploring more of the parameter space. MC3 algorithms explore and swap proposals with heated chains that continue to sample parameters from the prior distribution and from flattened likelihood surfaces. MC3 offers a relatively robust method for integrating marginal likelihoods [49].

### Variable selection

To assess whether the inclusion of migration between different communities is informative, and to establish if rates are seasonal, we implemented Bayesian variable selection [50] in an MC3 framework. Indicator variables $I_{Rfrom,to}$ and $I_{Sfrom,to}$ are added, giving the full parameterization of a single seasonal migration matrix cell as $I_{Rfrom,to} \cdot r_{from,to} \cdot (1 + I_{S\,from,to} \cdot \sigma_{from,to})$ and $I_{Rfrom,to} \cdot r_{from,to} \cdot (1 - I_{S\,from,to} \cdot \sigma_{from,to})$ for the other part of the year. In this case symmetric non-informative priors are used for the indicators. Reported Bayesian support for migration between two communities (BF) is the ratio of cases for samples for which an indicator variable $I_{Rfrom,to}$ is 1 vs. 0. Similarly this ratio is used to show support for seasonal migration between two communities.

### Combined likelihood based on the hemagglutinin and neuraminidase proteins

We use a conservative approach to combine the information present in both protein trees with respect to model likelihood. The combined protein tree log-likelihood is weighed down by half, to account for the possible lack of independence in the information contained in the two trees with respect to migration rates and seasonality. This choice does not affect the maximum likelihood model but has the effect of widening confidence intervals when the two trees provide independent data, while providing the correct confidence interval when the two proteins are in complete linkage.

### Availability of supporting data

The datasets supporting the results of this article are available online. The GenBank accession numbers of sequences used in this study and their geographic classification are available in Additional file 3. BEAST XML, tree and log files are available through the Dryad data repository under doi:10.5061/dryad.t120k.

### Additional files

**Additional file 1: Supplementary figures.**

**Additional file 2: Supplementary Methods for Bayesian Inference and Stochastic Mapping of Seasonal Migration Processes from Phylogenetic Tree Distributions** (SeasMig).

**Additional file 3: Genbank accession numbers.**

**Competing interests**
The authors declare that they have no competing interests.

**Authors' contributions**
DZ: conceived the project. DZ, TB, EB and RW added methodological contributions DZ, TB, RW, MR and MP analyzed the results and contributed to drafting and writing of the paper. All authors read and approved the final manuscript.

**Acknowledgments**
We thank Stephen Smith for comments and discussion. MP is an investigator of the Howard Hugues Medical Institue.

**Author details**
[1]Department of Computational Medicine and Bioinformatics, University of Michigan, Ann-Arbor, MI 48109, USA. [2]Vaccine and Infectious Disease Division, Fred Hutchinson Cancer Research Center, Seattle, WA 98109, USA. [3]Department of Ecology and Evolutionary Biology, University of Michigan, Ann-Arbor, MI 48109, USA. [4]Department of Internal Medicine, Division of Infectious Diseases, University of Michigan Health System, Ann-Arbor, MI 48109, USA. [5]Howard Hughes Medical Institute, Ann-Arbor, MI 48109, USA.